\documentclass[prd,twocolumn,showpacs,floatfix,amsmath,nofootinbib,amssymb,floatfix]{revtex4}
\usepackage{graphicx,color,dcolumn,booktabs,bm}
\usepackage{longtable,lscape}
\usepackage{txfonts}
\usepackage{overpic}
\usepackage{amssymb}
\usepackage{indentfirst}
\usepackage{feynmf}   %{feynmp}
\usepackage{slashed}  %for Feynman symbols
\usepackage{cases}
\usepackage{color}
\usepackage{multirow}
\usepackage{epstopdf}
\usepackage{longtable}
\usepackage{float}
\usepackage{graphicx,color,dcolumn,booktabs,bm}
\usepackage[colorlinks,
            citecolor=blue,
            anchorcolor=red,
            menucolor=red,
            linkcolor=red,
            filecolor=red,
            runcolor=red,
            urlcolor=blue,
            frenchlinks=red]{hyperref}

\graphicspath{{Figures}} %

\allowdisplaybreaks

\begin{document}
%\begin{CJK}{GBK}{}

\title{$D^{(*)}\Delta$ interactions and the molecular interpretation of $\Sigma_c(3200)^0$}

\author{Rui Chen$^{1,2,4}$}\email{chenrui@hunnu.edu.cn}
\author{Yu-Yue Cui$^{1}$}
\author{Fu-Lai Wang$^{3,4}$}

\affiliation{
$^1$Key Laboratory of Low-Dimensional Quantum Structures and Quantum Control of Ministry of Education, Department of Physics and Synergetic Innovation Center for Quantum Effects and Applications, Hunan Normal University, Changsha 410081, China\\
$^2$Hunan Research Center of the Basic Discipline for Quantum Effects and Quantum Technologies, Hunan Normal University, Changsha 410081, China\\
$^3$School of Physical Science and Technology, Lanzhou University, Lanzhou 730000, China\\
$^4$Lanzhou Center for Theoretical Physics, Lanzhou University, Lanzhou 730000, China
}
\date{\today}

\begin{abstract}

Inspired by the recent observation of the $\Sigma_c(3200)^0$ baryon in the $\Lambda_c^+\pi^-$ invariant mass spectrum by the LHCb Collaboration, we perform a comprehensive investigation of the $D^{(*)}\Delta$ interactions within the one-boson-exchange framework. Our results show that the $D^*\Delta$ molecule with $I(J^P)=1(1/2^-)$ can be regarded as the most natural candidate for the $\Sigma_c(3200)$, and predict the existence of several $D^*\Delta$ molecular partners with $I(J^P)=1(3/2^-, 5/2^-)$ and $2(1/2^{\pm}, 3/2^{\pm}, 5/2^-)$. After including the coupled channel effects between $D\Delta$ and $D^*\Delta$, several shallowly bound $D\Delta$-dominated molecular candidates also emerge, especially for the $I(J^P)=1,2(3/2^-)$ and the $2(1/2^+,3/2^+)$ systems. These results provide a unified molecular interpretation of the newly observed $\Sigma_c(3200)^0$ and useful guidance for future experimental searches for exotic charmed-baryon states. We expect the high-statistics data from the LHCb Upgrade and Belle~II runs will directly test these predictions.

\end{abstract}

%\keywords{Molecular State, One-boson-exchange model}

\maketitle

\section{Introduction}

Quantum chromodynamics (QCD), as the fundamental theory of strong interactions, exhibits non-perturbative behavior at low energies, making the understanding of hadron structure and hadron–hadron interactions from first principles one of the most challenging subjects in particle physics. The conventional quark model classifies hadrons as mesons (composed of a quark–antiquark pair) and baryons (three quarks). However, QCD itself does not exclude the existence of more complex hadronic configurations, such as glueballs, hybrid mesons, multiquark states, and hadronic molecules, which are generically referred to as exotic hadrons. Over the past two decades, a large number of new hadronic states beyond the predictions of the conventional quark model have been discovered by various experimental facilities (e.g., BaBar, Belle, and LHCb), providing unprecedented opportunities for studying hadronic structure in the non-perturbative regime of QCD~\cite{Bai:2026atm,Wang:2025dur,Liu:2013waa, Hosaka:2016pey, Chen:2016qju, Richard:2016eis, Lebed:2016hpi, Brambilla:2019esw, Liu:2019zoy, Chen:2022asf, Olsen:2017bmm, Guo:2017jvc, Meng:2022ozq}.

Among these exotic candidates, hadronic molecules, loosely bound states of two or more color-singlet hadrons via residual strong interactions, have attracted considerable interest due to their similarity to the physics of atomic nuclei. Typical hadronic molecules lie near the thresholds of their constituent hadrons, with small binding energies and large spatial sizes. The famous $X(3872)$ is widely considered as a strong candidate for a $D\bar D^*$ molecular state~\cite{Wang:2024ytk,Fleming:2007rp,Tornqvist:2004qy,Swanson:2003tb,Braaten:2003he,Voloshin:2003nt}, while the hidden-charm pentaquark states $P_c(4312)$, $P_c(4440)$, and $P_c(4457)$ discovered by the LHCb Collaboration~\cite{LHCb:2019kea} are often interpreted as molecular states of anticharmed mesons and charmed baryons~\cite{Yang:2011wz,Wang:2011rga,Karliner:2015ina,Wu:2010jy,Chen:2019asm,Meng:2019ilv,Liu:2019tjn,Xiao:2019aya,Guo:2019fdo,Xiao:2019mvs,
Guo:2019kdc,Burns:2019iih,Zhu:2019iwm,Chen:2020uif,Wang:2020eep,Chen:2020kco,Liu:2020hcv,Zhu:2021lhd,Xiao:2021rgp,Wang:2022neq,Meng:2022wgl,Zhu:2022wpi,Feijoo:2022rxf}. These findings indicate that hadronic molecules containing charm quarks serve as an important platform for studying the low-energy dynamics of QCD.

In parallel with the hidden-charm pentaquark states, significant efforts have been devoted to understanding ordinary charmed baryons, particularly the excited $\Lambda_c$ and $\Sigma_c$ states, as hadronic molecules composed of a $D^{(*)}$ meson and a nucleon~\cite{He:2006is,He:2010zq,Ortega:2012cx,Dong:2009tg,Dong:2010gu,Dong:2010xv,Wang:2020dhf,Lutz:2005ip,Wang:2018jaj,Sakai:2020psu,Xin:2023gkf}. This scenario is theoretically attractive because the $DN$ and $D^*N$ thresholds lie in the mass region around 2.8--3.0 GeV, where several excited charmed baryons are experimentally observed~\cite{Belle:2004zjl,BaBar:2006itc,LHCb:2017jym,Belle:2022hnm}. The $\Lambda_c(2595)$ and $\Lambda_c(2625)$ are among the earliest candidates for such a molecular interpretation~\cite{Romanets:2012hm,Liang:2014kra}. In the framework of unitarized coupled-channel approaches with chiral dynamics, these two low-lying excited $\Lambda_c$ states can be dynamically generated from the $DN$ and $D^*N$ interactions with coupled channels~\cite{Romanets:2012hm,Liang:2014kra}. Specifically, $\Lambda_c(2595)$, with $J^P = 1/2^-$, couples predominantly to $DN$ and $D^*N$, while $\Lambda_c(2625)$, with $J^P = 3/2^-$, couples mainly to $D^*N$. This dynamical generation mechanism, analogous to the $\Lambda(1405)$ as a $\bar K N$ quasi-bound state in the strangeness sector, provides a natural explanation for the properties of these resonances~\cite{Hyodo:2011ur,Ezoe:2016mkp,Hyodo:2007jq,Jido:2003cb,Oset:1997it}. Subsequent studies of semileptonic $\Lambda_b$ decays have further supported the molecular picture of these two $\Lambda_c^*$ resonances~\cite{Liang:2016ydj,Liang:2016exm}.

At higher masses, the $\Sigma_c(2800)$ and the $\Lambda_c(2940)^+$ have attracted particular attention as potential $S$-wave $DN$ and $D^*N$ molecular states, respectively~\cite{Dong:2010gu,Wang:2020dhf,Sakai:2020psu,Zhang:2012jk,Xin:2023gkf}. Dong et al.~proposed that $\Sigma_c(2800)$ could be interpreted as a $ND$ hadronic molecule and computed its strong decay widths to $\Lambda_c\pi$, finding consistency with experimental data for the $J^P = 1/2^+$ or $3/2^-$ assignments~\cite{Dong:2010gu}. In the chiral effective field theory approach to next-to-leading order, Wang et al.~studied the $DN$ and $D^*N$ interactions and obtained masses of $2792$ MeV for the isoscalar $[DN]_{J=1/2}^{I=0}$ state and around $2940$ MeV for the isoscalar $[D^*N]$ molecular states, suggesting that $\Lambda_c(2940)$ is likely an isoscalar $D^*N$ molecule~\cite{Wang:2020dhf}. QCD sum rule calculations have also been employed to examine the possibility of $\Sigma_c(2800)$ and $\Lambda_c(2940)^+$ as $S$-wave $D^{(*)}N$ molecular states~\cite{Zhang:2012jk,Xin:2023gkf}. However, the interpretation of these states remains controversial. A recent quark model study using the quark delocalization color screening model found that while $\Lambda_c(2940)$ can be interpreted as a $J^P = 3/2^-$ molecular state with a dominant $ND^*$ component, $\Sigma_c(2800)$ cannot be explained as the $ND$ molecular state with $J^P = 1/2^-$~\cite{Yan:2022nxp}. Moreover, $\Lambda_c(2910)$ recently observed by Belle~\cite{Belle:2022hnm} is unlikely to be a molecular state in the same framework~\cite{Yan:2022nxp}.

Beyond the $DN$ and $D^*N$ systems, the extension to the $D^*\Delta$ channel is a natural next step. Given that the $\Delta$ resonance is the spin-3/2 excitation of the nucleon, the $D^{(*)}\Delta$ systems offer new quantum number combinations and thresholds that may accommodate additional exotic charmed baryons. In fact, the BaBar collaboration observed a structure $X_c(3250)$ in the $\Sigma_c^{++}\pi^-\pi^-$ invariant mass spectrum with mass $M = 3245 \pm 20\,\text{MeV}/c^2$ and width $\Gamma = 108 \pm 60\,\text{MeV}/c^2$~\cite{BaBar:2012ekl}, which was already proposed as a possible $D^*\Delta$ molecular candidate~\cite{Ortega:2014fha}. Within the quark model framework, it has been shown that the $D^*\Delta$ system can form three molecular states with different quantum numbers: $I(J^P) = 2(1/2^-)$, $2(3/2^-)$, and $1(5/2^-)$, with masses compatible with the experimental value of $X_c(3250)$~\cite{Ortega:2014fha,Zhang:2020dwp,Carames:2012bd,Zhao:2016zhf}. Furthermore, studies of the $D^{(*)}\Delta $ system in the chiral SU(3) quark model indicate that the system can form several bound states with attractive interactions~\cite{Zhang:2020dwp}.

Very recently, the LHCb Collaboration performed an amplitude analysis of the decay $B^- \to \Lambda_c^+ \bar p \pi^-$ using proton--proton collision data corresponding to an integrated luminosity of $9\,\text{fb}^{-1}$ at center-of-mass energies of 7, 8, and 13 TeV~\cite{LHCb:2026nzu}. In the invariant mass spectrum of $\Lambda_c^+ \pi^-$, in addition to the known $\Sigma_c(2455)^0$, $\Sigma_c(2520)^0$, and $\Sigma_c(2800)^0$ baryons, two new states were observed: $\Sigma_c(2900)^0$ and $\Sigma_c(3200)^0$. The masses and widths of these higher-mass states were determined using Breit--Wigner parameterizations. This discovery significantly enriches charmed baryon spectroscopy and provides key data for understanding the structure and dynamics of charmed baryons. It is noteworthy that the mass of $\Sigma_c(3200)^0$ is very close to the $D^*\Delta$ threshold (about $3240\,\text{MeV}$). This feature strongly suggests that $\Sigma_c(3200)^0$ may have a $D^*\Delta$ molecular nature.

However, a systematic study of the interaction between ground-state charmed mesons ($D, D^*$) and the $\Delta$ baryon is still lacking. In particular, questions remain regarding whether $\Sigma_c(3200)^0$ can indeed be interpreted as a $D^*\Delta$ molecule, and how its quantum numbers can be determined. In this work, we aim to systematically investigate the interactions between ground-state charmed mesons and the $\Delta$ baryon, and to explore whether the newly discovered $\Sigma_c(3200)^0$ by the LHCb Collaboration can be interpreted as a $D^*\Delta$ hadronic molecule. Meanwhile, we can predict the existence of possible charm baryon-like molecules. 

This paper is organized as follows. In Sec.~\ref{sec2}, we derive the OBE effective potentials for the $D^{(*)}\Delta$ systems. The numerical results are presented and discussed in Sec.~\ref{sec3}. Finally, a summary is provided in Sec.~\ref{sec4}.

\section{The OBE interactions}\label{sec2}

Before deducing the OBE effective potentials for the $D^{(*)}\Delta$ systems, we first construct the wave functions, which include the isospin wave functions, the spin-orbit wave functions, and the radial wave functions. The flavor wave functions $|I,I_3\rangle$ for all the investigated systems can be expressed as
\begin{eqnarray*}
|1,1\rangle &= &\frac{1}{2}(D^{(*)+}\Delta^++\sqrt{3}{D}^{(*)0}\Delta^{++}
),\\
|1,0\rangle &= &\frac{1}{\sqrt{2}}(D^{(*)+}\Delta^0+{D}^{(*)0}\Delta^{+}
),
  \\
|1,-1\rangle&= &\frac{1}{2}(D^{(*)0}\Delta^0+\sqrt{3}{D}^{(*)+}\Delta^{-}
),   \\  
|2,2\rangle&= &D^{(*)+}\Delta^{++},   \\ 
|2,1\rangle&= &\frac{1}{2}({D}^{(*)0}\Delta^{++}-\sqrt{3}D^{(*)+}\Delta^+
), \\ 
|2,0\rangle&= &\frac{1}{\sqrt{2}}({D}^{(*)0}\Delta^{+}-D^{(*)+}\Delta^0
),\\
|2,-1\rangle&= &\frac{1}{2}(\sqrt{3}{D}^{(*)0}\Delta^{0}-D^{(*)+}\Delta^-
),\\
|2,-2\rangle&= &{D}^{(*)0}\Delta^{-}.
\end{eqnarray*}
When we study the $S$-$D$ wave mixing and $P$-wave interactions, the spin-parities $J^{P}$ for the $D^{(*)}\Delta$ systems include $1/2^{\pm}$, $3/2^{\pm}$, $5/2^{\pm}$, and $7/2^+$. In Table \ref{channels}, we collect the spin-orbit wave functions $|{}^{2S+1}L_J\rangle$ for the channels with different quantum number configurations, where the spin-orbit wave functions have the general form as
\begin{eqnarray}
|D\Delta({}^{2S+1}L_{J})\rangle &=& \sum_{m,m_L}C^{J,M}_{\frac{3}{2},Lm_L}
          \Phi_{\frac{3}{2}m}|Y_{L,m_L}\rangle,\label{001}\nonumber\\
|D^*\Delta({}^{2S+1}L_{J})\rangle &=& \sum_{m,m',m_Sm_L}C^{S,m_S}_{\frac{3}{2}m,1m'}C^{J,M}_{Sm_S,Lm_L}
          \Phi_{\frac{3}{2}m}\epsilon_{m'}|Y_{L,m_L}\rangle,\label{002}\nonumber
\end{eqnarray}
Here, $C^{J,M}_{\frac{3}{2},Lm_L}$, $C^{S,m_S}_{\frac{3}{2}m,1m'}$, and $C^{J,M}_{Sm_S,Lm_L}$ are the Clebsch-Gordan coefficients. $Y_{L,m_L}$ is the spherical harmonic functions. $\epsilon$ and $\Phi_{\frac{3}{2}m}$ are the polarization vector for the ${D}^*$ vector meson and the polarization tensor for the $\Delta$ baryon, which have the forms of $\epsilon_{\pm}^{m}=\mp\frac{1}{\sqrt{2}}\left(\epsilon_x^{m}{\pm}i\epsilon_y^{m}\right)$ and $\epsilon_0^{m}=\epsilon_z^{m}$, and $\Phi_{\frac{3}{2}m}=\sum_{m_1,m_2}\langle\frac{1}{2},m_1;1,m_2|\frac{3}{2},m\rangle\chi_{\frac{1}{2} m_1}\epsilon^{m_2}$, these components satisfy $\epsilon_{\pm1}= \frac{1}{\sqrt{2}}\left(0,\pm1,i,0\right)$ and $\epsilon_{0} =\left(0,0,0,-1\right)$. 

\renewcommand\tabcolsep{0.47cm}
\renewcommand{\arraystretch}{1.7}
\begin{table}[!htpb]
\centering
\caption{The spin-orbit wave functions $|{}^{2S+1}L_{J}\rangle$ for the discussed $D^{(*)}\Delta$ systems.}\label{channels}
\begin{tabular}{ll|lll}\toprule[1.0pt]\midrule[1.0pt]
     $J^P$   &$D\Delta$   & $J^P$  &$D^*\Delta$ \\\midrule[1.0pt]
 $\frac{1}{2}^-$    &$| {}^4{D}_{\frac{1}{2}}\rangle$    &$\frac{1}{2}^-$    &$|{}^2{S}_{\frac{1}{2}}, {}^4{D}_{\frac{1}{2}}, {}^6{D}_{\frac{1}{2}}\rangle$\\
$\frac{3}{2}^-$    &$|{}^4{S}_{\frac{3}{2}}, {}^4{D}_{\frac{3}{2}}\rangle$  &$\frac{3}{2}^-$    &$|{}^4{S}_{\frac{3}{2}}, {}^2{D}_{\frac{3}{2}},  {}^4{D}_{\frac{3}{2}}, {}^6{D}_{\frac{3}{2}}\rangle$\\
$\frac{1}{2}^+$    &$| {}^4{P}_{\frac{1}{2}}\rangle$    &$\frac{5}{2}^-$    &$|{}^6{S}_{\frac{5}{2}}, {}^2{D}_{\frac{5}{2}},  {}^4{D}_{\frac{5}{2}}, {}^6{D}_{\frac{5}{2}}\rangle$\\  
$\frac{3}{2}^+$    &$| {}^4{P}_{\frac{3}{2}}\rangle$    &$\frac{1}{2}^+$    &$|{}^2{P}_{\frac{1}{2}}, {}^4{P}_{\frac{1}{2}}\rangle$\\ 
$\frac{5}{2}^+$    &$| {}^4{P}_{\frac{5}{2}}\rangle$    &$\frac{3}{2}^+$    &$|{}^2{P}_{\frac{3}{2}}, {}^4{P}_{\frac{3}{2}}, {}^6{P}_{\frac{3}{2}}\rangle$\\ 
 & &$\frac{5}{2}^+$    &$| {}^4{P}_{\frac{5}{2}}, {}^6{P}_{\frac{5}{2}}\rangle$ \\
  & &$\frac{7}{2}^+$    &$| {}^6{P}_{\frac{7}{2}}\rangle$ \\
\bottomrule[1.0pt]\midrule[1.0pt]
\end{tabular}
\end{table}

Utilizing the restrictions of the heavy quark symmetry, the chiral symmetry, and the hidden local symmetry \cite{Casalbuoni:1992gi,Casalbuoni:1996pg,Yan:1992gz,Wise:1992hn,Ding:2008gr}, one can construct the effective Lagrangians description of the interactions between the charmed mesons ${D}^{(*)}$ and the light mesons, which are explicitly written as
\begin{eqnarray}\label{lag1}
\mathcal{L}_{H}&=&g_{\sigma}\left\langle H^{(Q)}_a\sigma\bar{H}^{(Q)}_a\right\rangle+ig\left\langle H^{(Q)}_b{\mathcal A}\!\!\!\slash_{ba}\gamma_5\bar{H}^{\,({Q})}_a\right\rangle\nonumber\\
  &&+i\beta\left\langle H^{(Q)}_b v^{\mu}({\mathcal V}_{\mu}-\rho_{\mu})_{ba}\bar{H}^{\,(Q)}_a\right\rangle\nonumber\\
  &&+i\lambda\left\langle H^{(Q)}_b \sigma^{\mu\nu}F_{\mu\nu}(\rho)_{ba}\bar{H}^{\,(Q)}_a\right\rangle.
\end{eqnarray}
The superfield $H^{({Q})}_a$, constructed from the $S$-wave charmed mesons ${D}$ with $J^P=0^-$ and ${D}^{*}$ with $J^P=1^-$, is defined as
\begin{eqnarray}
H^{(Q)}_a&=&\frac{1+{v}\!\!\!\slash}{2}\left(D^{*\mu}_a\gamma_{\mu}-D_a\gamma_5\right).
\end{eqnarray} In addition, the associated conjugate field is  $\bar{H}^{(Q)}_a=\gamma^0H_a^{(Q)\dag}\gamma^0$. 
Here, the four-velocity is represented as $v_{\mu}$ and has the value $v_{\mu}=(1,\bm{0})$ when making the non-relativistic approximation. For the previously defined superfield  $H^{({Q})}_a$, the normalization relations for the charmed mesons $D$ and $D^{*}$ are $\langle 0|D|c\bar{q}\left(0^-\right)\rangle=\sqrt{m_{D}}$
and
$\langle 0|D^{*\mu}|c\bar{q}\left(1^-\right)\rangle=\sqrt{m_{D^*}}\epsilon^\mu$. $\mathcal{A}_\mu$ and ${\mathcal V}_{\mu}$ are the axial current and the vector current, which can be written as
\begin{eqnarray}
{\mathcal A}_{\mu}=\frac{1}{2}\left(\xi^{\dagger}\partial_{\mu}\xi-\xi\partial_{\mu}\xi^{\dagger}\right),~~
{\mathcal V}_{\mu}=\frac{1}{2}\left(\xi^{\dagger}\partial_{\mu}\xi+\xi\partial_{\mu}\xi^{\dagger}\right),
\end{eqnarray}
respectively. $\xi=e^{i\mathbb{P}/f_\pi}$ is the pseudo-Goldstone meson field with $f_{\pi}$ being equal to $132~{\rm MeV}$. Furthermore, the vector meson field and the vector meson field strength tensor are defined as $\rho_{\mu}$ and $F_{\mu\nu}(\rho)$, which are explicitly written as
$\rho_{\mu}=i{g_V}\mathbb{V}_{\mu}/{\sqrt{2}}$ and
$F_{\mu\nu}(\rho)=\partial_{\mu}\rho_{\nu}-\partial_{\nu}\rho_{\mu}+[\rho_{\mu},\rho_{\nu}]$,
respectively. 

Once expanding the Eq. (\ref{lag1}), we can obtain
\begin{eqnarray}
\mathcal{L}_{D^{(*)}D^{(*)} \sigma} &=&-2g_{\sigma}{D}_a \sigma {D}_a^{\dag}+ 2g_{\sigma} {D}_{a\mu}^* \sigma {D}_a^{*\mu\dag},\\
%%%
\mathcal{L}_{D^{(*)}D^{(*)}\mathbb{P}}&=&-\frac{2ig}{f_{\pi}}v^{\alpha}\varepsilon_{\alpha\mu\nu\lambda}D_b^{*\mu}D_a^{*\lambda\dag}\partial^{\nu}{\mathbb{P}}_{ba}\nonumber\\
    &&-\frac{2g}{f_{\pi}}(D_b^{*\mu}D_a^{\dag}+D_bD_a^{*\mu\dag})\partial_{\mu}{\mathbb{P}}_{ba},\\
%%%%%%
\mathcal{L}_{D^{(*)}D^{(*)}\mathbb{V}} &=&-\sqrt{2}\beta g_V D_b D_a^{\dag} v\cdot\mathbb{V}_{ba}+\sqrt{2}\beta g_V D_{b\mu}^* D_a^{*\mu\dag}v\cdot\mathbb{V}_{ba}\nonumber\\
    &&-2\sqrt{2}i\lambda g_V D_b^{*\mu}D_a^{*\nu\dag}\left(\partial_{\mu}\mathbb{V}_{\nu}-\partial_{\nu}\mathbb{V}_{\mu}\right)_{ba}\nonumber\\
    &&-2\sqrt{2}\lambda g_V v^{\lambda}\varepsilon_{\lambda\mu\alpha\beta}(D_bD_a^{*\mu\dag}+D_b^{*\mu}D_a^{\dag})\partial^{\alpha}\mathbb{V}^{\beta}_{ba}.
\end{eqnarray}
Here, $\mathbb{P}$ and $\mathbb{V}$ correspond to the pseudoscalar and vector mesons fields, respectively, which have the forms of 
\begin{eqnarray}
{\mathbb{P}} &=& {\left(\begin{array}{ccc}
       \frac{\pi^0}{\sqrt{2}}+\frac{\eta}{\sqrt{6}} &\pi^+ \\
       \pi^-       &-\frac{\pi^0}{\sqrt{2}}+\frac{\eta}{\sqrt{6}}      \end{array}\right)},\\
{\mathbb{V}}_{\mu} &=& {\left(\begin{array}{ccc}
       \frac{\rho^0}{\sqrt{2}}+\frac{\omega}{\sqrt{2}} &\rho^+ \\
       \rho^-       &-\frac{\rho^0}{\sqrt{2}}+\frac{\omega}{\sqrt{2}}      \end{array}\right)}_{\mu}.
\end{eqnarray}

For the interactions between the $\Delta$ baryons and the light mesons, we adopt the following expressions \cite{Matsuyama:2006rp,Ronchen:2012eg}, i.e.,
\begin{eqnarray}
\mathcal{L}_{\Delta\Delta\sigma} &  = & -g_{\Delta\Delta\sigma} \bar{\Delta}_{\mu} \sigma \Delta^{\mu}, \\
%%%%%%
\mathcal{L}_{\Delta\Delta\pi}&=& \frac{g_{\Delta\Delta\pi}}{m_{\pi}}~\bar{\Delta}_{\mu}\gamma^5\gamma^{\nu}{\bm T}\cdot\partial_{\nu}{\bm \pi}\Delta^{\mu},\\
%%%%%%
\mathcal{L}_{\Delta\Delta\mathbb{\rho}}&=& -g_{\Delta\Delta\mathbb{\rho}}~\bar{\Delta}_{\tau}(\gamma^{\mu}-\frac{\kappa_{\Delta\Delta\mathbb{\rho}}}{2m_{\Delta}}\sigma^{\mu\nu}\partial_{\nu}){ \mathbb{V}}_{\mu}\cdot {\bm T}\Delta^{\tau},\\
%%%%%%
\mathcal{L}_{\Delta\Delta\mathbb{\omega}}&=& -g_{\Delta\Delta\mathbb{\omega}}~\bar{\Delta}_{\tau}(\gamma^{\mu}-\frac{\kappa_{\Delta\Delta\mathbb{\omega}}}{2m_{\Delta}}\sigma^{\mu\nu}\partial_{\nu}){ \mathbb{V}}_{\mu}\Delta^{\tau}.
\end{eqnarray}
In above Lagrangians, $\bm{T}\cdot\varphi$ with $\varphi=\pi, \rho$ is expressed as
\begin{eqnarray}
{\bm T}\cdot{\bm\varphi}&=&\sqrt{\frac{4}{15}}\left(\begin{array}{cccc}
\frac{3}{2}\varphi^0 & \sqrt{\frac{3}{2}}\varphi^+& 0& 0\\
\sqrt{\frac{3}{2}}\varphi^- & \frac{1}{2}\varphi^0 & \sqrt{2}\varphi^+& 0\\
0 & \sqrt{2}\varphi^- & -\frac{1}{2}\varphi^0 &\sqrt{\frac{3}{2}}\varphi^+
\\ 0 & 0 &\sqrt{\frac{3}{2}}\varphi^- &-\frac{3}{2}\varphi^0
\end{array}\right),
\end{eqnarray}
where notations $+$, $-$, and $0$ denote the charges of the mesons.

The coupling constants $g^2_{\Delta\Delta\sigma}/4\pi=5.69$,  $g_{\Delta\Delta\pi}=1.78$, $g_{\Delta\Delta\rho}=4.9$, and $\kappa_{\Delta\Delta\rho}=6.1$ were obtained from fitting the experimental data in Refs.~\cite{Sato:1996gk,Matsuyama:2006rp,Ronchen:2012eg}. The coupling constants for the $\omega$ meson can be related to these for the $\rho$ meson with SU(3) symmetry as  $g_{\Delta\Delta\omega}=3/2g_{\Delta\Delta\rho}$, and $\kappa_{\Delta\Delta\omega}=\kappa_{\Delta\Delta\rho}$. In addition, the phase factors between the relevant coupling constants can then be determined by applying the quark model \cite{Riska:2000gd}. In the present work, the values of $g_{\sigma}=2.82$, $g=-0.62$, $\beta g_{V}=2.53$, and $\lambda g_{V}=3.65~\rm {GeV^{-1}}$ are taken from Refs. \cite{Wang:2019aoc, Wang:2011rga, Machleidt:2000ge, Machleidt:1987hj, Cao:2010km}. 

After prepared these effective Lagrangians, we next derive the scattering amplitudes $\mathcal{M}$ for $D^{(*)}\Delta\to D^{(*)}\Delta$ processes in t-channel by exchanging one light meson. The OBE effective potentials for the discussed channels in momentum space can be obtained with the help of the Breit approximation, i.e.,
\begin{eqnarray}
  \mathcal{V}^{\mathbb{E}}(\bm{q}) = -\frac{\mathcal{M}(D^{(*)}\Delta\to D^{(*)}\Delta)}{\sqrt{2m_{D^{(*)}}^i 2m_{D^{(*)}}^f 2m_{\Delta}^i 2m_{\Delta}^f}}.
  \end{eqnarray}
Here, notations $i$ and $f$ label the initial and final states, respectively. $\mathbb{E}$ stands for the exchanged mesons, including the scalar meson $\sigma$, the pseudoscalar mesons $\pi/\eta$, the vector mesons $\rho/\omega$. After performing the Fourier transform, we can finally get the OBE effective potentials in the coordinate space, i.e.,
\begin{eqnarray}
\mathcal{V}^{\mathbb{E}}(r) = \int \frac{d^3\bm{q}}{(2\pi)^{3}}e^{i\bm{q}\cdot \bm{r}}\mathcal{V}^{\mathbb{E}}(\bm{q})\mathcal{F}^2(q^2,m_{\mathbb{E}}^2).
\end{eqnarray}
Here, a monopole form factor, $\mathcal{F}(q^2,m_{\mathbb{E}}^2)=(\Lambda^2-m_{\mathbb{E}}^2)/(\Lambda^2-q^2)$, is introduced at every interaction vertexes, which can compensate the off-shell effects of the exchanged particles. $\Lambda$, $m_{\mathbb{E}}$, and $q$ denote the cutoff, the mass, and the four momentum of the exchanged mesons. The reasonable value of the cutoff $\Lambda$ are in the range of $\Lambda\sim1$ GeV, based on the experience of the study of deuteron~\cite{Tornqvist:1993ng,Tornqvist:1993vu}.

With these procedures, we finally deduce the OBE effective potentials for the $D^{(*)}\Delta$ systems, i.e.,
\begin{align}
\mathcal{V}_{D^{(*)}\Delta} &=&\mathcal{V}_{D^{(*)}\Delta}^{\sigma}+\mathcal{G}(I)\mathcal{V}_{D^{(*)}\Delta}^{\pi}+\mathcal{G}(I)\mathcal{V}_{D^{(*)}\Delta}^{\rho}+\mathcal{V}_{D^{(*)}\Delta}^{\omega}.
\end{align}
where $\mathcal{G}(I)$ is the isospin factor: $\mathcal{G}(I=1)=\frac{\sqrt{30}}{6}$ and $\mathcal{G}(I=2)=-\frac{\sqrt{30}}{10}$. And $\mathcal{V}_{D^{(*)}\Delta}^{{\mathbb{E}}}$ are the OBE potentials for the process $D^{(*)}\Delta\to D^{(*)}\Delta$ by exchanging meson $\mathbb{E}$, i.e.,
\begin{eqnarray}
\mathcal{V}_{D\Delta\to D\Delta}^{\sigma} &=&C_1 \mathcal{A}_1 Y_{\Lambda,m_\sigma},\\
\mathcal{V}_{D\Delta\to D\Delta}^{\rho/\omega}&=&-\frac{\sqrt{2}C_3}{2}\mathcal{A}_1Y_{\Lambda,m_V}\\
\mathcal{V}_{D^*\Delta\to  D^*\Delta}^\sigma &=&C_1\mathcal{A}_2 Y_{\Lambda,m_\sigma},\\
\mathcal{V}_{D^*\Delta\to    D^*\Delta}^{\pi} &=&
\frac{C_2}{3}(\mathcal{A}_3 Z_{\Lambda,m_{\pi}} 
 +\mathcal{A}_4 T_{\Lambda,m_{\pi}}), 
  \label{pot1}\\
\mathcal{V}_{D^*\Delta\to D^*\Delta}^{\rho,\omega} 
  &=&-\frac{\sqrt{2}C_3}{6}(1+k_{\Delta\Delta V})(2\mathcal{A}_3 Z_{\Lambda,m_V}-\mathcal{A}_4 T_{\Lambda,m_V})\nonumber\\
  &&
  -\frac{C_4}{\sqrt{2}} \mathcal{A}_2 Y_{\Lambda,m_V},\\
\mathcal{V}_{D\Delta\to D^*\Delta}^{\pi} &=&
-\frac{ C_2}{3}(\mathcal{A}_5Z_{\Lambda,m_P}+\mathcal{A}_6 T_{\Lambda,m_P}),\\
\mathcal{V}_{D\Delta\to D^*\Delta}^{\rho,\omega} &=&\frac{\sqrt{2}C_3}{6}(1+k_{\Delta\Delta
\rho})(2\mathcal{A}_5Z_{\Lambda,m_P}-\mathcal{A}_6T_{\Lambda,m_P}).\quad
\end{eqnarray}
Here, we define several coupling constants, operators, and useful functions, whose explicit forms are 
\begin{align}
C_1 &= g_{\sigma}g_{ \Delta\Delta\sigma}, \quad 
C_3 = {g_V\beta g_{\Delta\Delta V}}, \\
C_2 &= \frac{g_{\Delta\Delta \pi}g}{f_{\pi} m_{\pi}}, \quad
C_4 = \frac{\lambda g_v g_{\Delta\Delta V}}{m_{\Delta}},\\
\mathcal{A}_1&=\sum_{m,n,a,b}C^{\frac{3}{2},m+n}_{1m,\frac{1}{2}n}C^{\frac{3}{2},a+b}_{1a,\frac{1}{2}b}\chi_4^{n\dagger} \chi^{\dagger}_3 \bm{\epsilon}^{m\dagger}_4\cdot \bm{\epsilon}_2^a \chi_2^b \chi_1,\label{operator1}\\
\mathcal{A}_2&=\sum_{m,n,a,b}C^{\frac{3}{2},m+n}_{1m,\frac{1}{2}n}C^{\frac{3}{2},a+b}_{1a,\frac{1}{2}b}\chi_4^{n\dagger}\chi_3^\dagger(\epsilon_1\cdot\epsilon_3^{\dagger})(\epsilon_2^a\cdot\epsilon_4^{m\dagger})\chi_2^b\chi_1,\\
\mathcal{A}_3&=\sum_{m,n,a,b}C^{\frac{3}{2},m+n}_{1m,\frac{1}{2}n}C^{\frac{3}{2},a+b}_{1a,\frac{1}{2}b}\chi_4^{n\dagger}\chi_3^\dagger(\bm\epsilon_2^a\cdot\bm\epsilon_4^{m\dagger})(i\bm\epsilon_1\times\bm\epsilon_3^{\dagger})\cdot\bm\sigma \chi_2^b\chi_1,\\
\mathcal{A}_4&=\sum_{m,n,a,b}C^{\frac{3}{2},m+n}_{1m,\frac{1}{2}n}C^{\frac{3}{2},a+b}_{1a,\frac{1}{2}b}\chi_4^{n\dagger}\chi_3^\dagger(\bm\epsilon_2^a\cdot\bm\epsilon_4^{m\dagger})\nonumber\\
  &\quad\quad\times S({i\bm\epsilon_1\times\bm\epsilon_3^{\dagger}},\bm\sigma,\hat{\bm{r}})\chi_2^b\chi_1,\\
\mathcal{A}_5&=\sum_{m,n,a,b}C^{\frac{3}{2},m+n}_{1m,\frac{1}{2}n}C^{\frac{3}{2},a+b}_{1a,\frac{1}{2}b}\chi_4^{n\dagger}\chi_3^\dagger(\bm\epsilon_2^a\cdot\bm\epsilon_4^{m\dagger})(\bm\epsilon_3^{\dagger}\cdot\bm\sigma) \chi_2^b\chi_1,\\
\mathcal{A}_6&=\sum_{m,n,a,b}C^{\frac{3}{2},m+n}_{1m,\frac{1}{2}n}C^{\frac{3}{2},a+b}_{1a,\frac{1}{2}b}\chi_4^{n\dagger}\chi_3^\dagger(\bm\epsilon_2^a\cdot\bm\epsilon_4^{m\dagger})S({\bm\epsilon_3^{\dagger}},\bm\sigma,\hat{\bm{r}})\chi_2^b\chi_1,\label{operator2}\\
Y_{\Lambda,m}&=\frac{1}{4\pi r}(e^{-mr}-e^{-\Lambda r})-\frac{\Lambda^2-m^2}{8 \pi\Lambda}e^{-\Lambda r},\\
T_{\Lambda,m}&=r\frac{\partial}{\partial r}\frac{1}{r}\frac{\partial}{\partial r}Y_{\Lambda,m},\quad\quad
Z_{\Lambda,m}=\nabla^2Y_{\Lambda,m}.
\end{align}
When performing the numerical calculations, the operators in Eqs. (\ref{operator1})-(\ref{operator2}) will be replaced by numerical matrices, which are summarized in Table \ref{operator_transposed}.

\renewcommand\tabcolsep{0.25cm}
\renewcommand{\arraystretch}{1.7}
\begin{table}[!htbp]
\caption{Matrix elements for the spin-spin interactions and tensor force operators in the OBE effective potentials. Here, $\langle\mathcal{A}_{1}\rangle=\langle\mathcal{A}_{2}\rangle=\mathcal{I}$ with $\mathcal{I}$ being the identity matrix.}\label{operator_transposed}
{\begin{tabular}{c|cccccc}
\toprule[1pt]
\toprule[1pt]
$J^P$ &$\mathcal{A}_{3}$  &$\mathcal{A}_{4}$   \\\hline

$\frac{1}{2}^-$ &
$\begin{pmatrix}-\frac{5}{3} & 0 & 0\\ 0 & -\frac{2}{3} & 0\\ 0 & 0 & 1 \end{pmatrix}$
&$\begin{pmatrix}0 & -\frac{7}{3\sqrt{5}} & -\frac{2}{\sqrt{5}}\\ -\frac{7}{3\sqrt{5}} & -\frac{16}{15} & -\frac{1}{5}\\ -\frac{2}{\sqrt{5}} &  -\frac{1}{5} & -\frac{8}{5} \end{pmatrix}$ 
  \\
$\frac{3}{2}^-$ &$\begin{pmatrix}  -\frac{2}{3} & 0& 0& 0\\ 0  & -\frac{5}{3} & 0& 0\\ 0  & 0 & -\frac{2}{3}& 0\\ 0  & 0 & 0& 1\end{pmatrix}$ 
&$\begin{pmatrix}  0 & \frac{7}{3\sqrt{10}}& \frac{16}{15}& \frac{-1}{5}\sqrt{\frac{7}{2}}\\ \frac{7}{3\sqrt{10}}  & 0 & \frac{-7}{3\sqrt{10}}& \frac{2}{\sqrt{35}}\\  \frac{16}{15}  & \frac{-7}{3\sqrt{10}} & 0& \frac{-1}{\sqrt{14}}\\ \frac{-1}{5}\sqrt{\frac{7}{2}}  & \frac{2}{\sqrt{35}} & \frac{-1}{\sqrt{14}}& -\frac{4}{7}\end{pmatrix}$\\
$\frac{5}{2}^-$ &$\begin{pmatrix}  1 & 0& 0& 0\\ 0  & -\frac{5}{3} & 0& 0\\ 0  & 0 & -\frac{2}{3}& 0\\ 0  & 0 & 0& 1\end{pmatrix}$ 
&$\begin{pmatrix}  0 & \frac{-2}{\sqrt{15}}& \frac{1}{5}\sqrt{\frac{7}{3}}& \frac{2\sqrt{14}}{5}\\ \frac{-2}{\sqrt{15}}  & 0 & \frac{1}{3}\sqrt{\frac{7}{5}}& 4\sqrt{\frac{2}{105}}\\  \frac{1}{5}\sqrt{\frac{7}{3}}  & \frac{1}{3}\sqrt{\frac{7}{5}} & \frac{16}{21}& \frac{-1}{7}\sqrt{\frac{2}{3}}\\ \frac{2\sqrt{14}}{5}  & 4\sqrt{\frac{2}{105}} & \frac{-1}{7}\sqrt{\frac{2}{3}}& \frac{4}{7}\end{pmatrix}$\\
$\frac{1}{2}^+$ &
$\begin{pmatrix}-\frac{5}{3} & 0 \\ 0 & -\frac{2}{3}  \end{pmatrix}$
&$\begin{pmatrix}0 & -\frac{7}{3\sqrt{5}} \\ -\frac{7}{3\sqrt{5}} & -\frac{16}{15}  \end{pmatrix}$ 
  \\
  $\frac{3}{2}^+$ &
$\begin{pmatrix}-\frac{5}{3} & 0 & 0\\ 0 & -\frac{2}{3} & 0\\ 0 & 0 & 1 \end{pmatrix}$
&$\begin{pmatrix}0 & \frac{7}{15\sqrt{2}} & \frac{2\sqrt{3}}{5}\\ \frac{7}{15\sqrt{2}} & \frac{64}{75} & -\frac{7}{25}\sqrt{\frac{3}{2}}\\ \frac{2\sqrt{3}}{5} & -\frac{7}{25}\sqrt{\frac{3}{2}} & -\frac{28}{25} \end{pmatrix}$
  \\
  $\frac{5}{2}^+$ &
$\begin{pmatrix}-\frac{2}{3} & 0 \\ 0 &1  \end{pmatrix}$
&$\begin{pmatrix}\frac{-16}{75} & \frac{\sqrt{21}}{25} \\ \frac{\sqrt{21}}{25} & \frac{32}{25}  \end{pmatrix}$ 
  \\
    $\frac{7}{2}^+$ &
$\begin{pmatrix} 1  \end{pmatrix}$
&$\begin{pmatrix}\frac{-2}{5}   \end{pmatrix}$ 
  \\\bottomrule[1pt]
  $J^P$ &$\mathcal{A}_{5}$  &$\mathcal{A}_{6}$   \\\hline

$\frac{3}{2}^-$ &
$\begin{pmatrix}-\sqrt{\frac{5}{3}} & 0 & 0& 0\\ 0 &0 &-\sqrt{\frac{5}{3}}&0 \end{pmatrix}$ &
$\begin{pmatrix}0 & -\frac{1}{\sqrt{6}} & -\frac{4}{\sqrt{15}}& -\sqrt{\frac{21}{10}}\\ -\frac{4}{\sqrt{15}} &\frac{1}{\sqrt{6}} &0&-\sqrt{\frac{15}{14}} \end{pmatrix}$ 
  \\
  $\frac{1}{2}^+$ &
$\begin{pmatrix}0 & -\sqrt{\frac{5}{3}} \end{pmatrix}$ &
$\begin{pmatrix}\frac{1}{\sqrt{3}} & \frac{4}{\sqrt{15}}  \end{pmatrix}$ 
  \\
    $\frac{3}{2}^+$ &
$\begin{pmatrix}0 & -\sqrt{\frac{5}{3}} &0\end{pmatrix}$ &
$\begin{pmatrix}-\frac{1}{\sqrt{30}} & -\frac{16}{5\sqrt{15}} &-\frac{21}{5\sqrt{10}} \end{pmatrix}$ 
  \\
    $\frac{5}{2}^+$ &
$\begin{pmatrix} -\sqrt{\frac{5}{3}} &0\end{pmatrix}$ &
$\begin{pmatrix}\frac{4}{5\sqrt{15}} & \frac{3}{5}\sqrt{\frac{7}{5}}  \end{pmatrix}$ 
  \\
\bottomrule[1pt]
\bottomrule[1pt]
\end{tabular}}
\end{table}

\section{Numerical results}\label{sec3}

With these OBE effective potentials, we next solve the coupled Schr\"{o}dinger equations to search for reasonable bound states solutions, where the cutoff value varies from 0.78 GeV to 2.50 GeV. Based on previous studies of the deuteron and similar molecular states, a typical cutoff scale $\Lambda \sim 1$~GeV is known to characterize loosely bound hadronic molecules~\cite{Tornqvist:1993ng, Tornqvist:1993vu, Chen:2015loa, Liu:2011xc, Chen:2016ypj, Chen:2021vhg, Chen:2020kco, Chen:2022dad, Chen:2022onm, Wang:2020bjt, Wang:2019nwt, Chen:2019asm, Chen:2019uvv, He:2013nwa, Li:2012ss, Li:2012bt, Sun:2012sy, Sun:2011uh, Yang:2011wz, Lee:2011rka, Liu:2009ei, Liu:2008xz, Chen:2024xlw, Liu:2008fh, Yang:2021sue, Liu:2010xh, Qian:2024joy, Yamaguchi:2019seo, He:2019ify, Burns:2019iih, He:2011ed, Thomas:2008ja, Liu:2008tn, Lee:2009hy, Wu:2010jy, Huo:2024eew,Cui:2025elw,Tang:2025bcc}. The binding energy $E$ is typically of the order of a few to several tens of MeV, and the root-mean-square (RMS) radius $r_{\mathrm{RMS}}$ is around 1~fm or much larger. The partial-wave probability distribution further reveals the dominant channel and the importance of coupled-channel effects. Adopting these criteria, we systematically analyse the numerical results for both the single channel $D^*\Delta$ systems and the coupled channel $D^{(*)}\Delta$ systems. In this work, we first explore the possibility of the newly $\Sigma_c(3200)$ as an isovector $D^*\Delta$ molecule, and then we can predict other possible molecular candidates mainly composed of the $D^{(*)}\Delta$ states.

\subsection{$\Sigma_c(3200)$ as an isovector $D^*\Delta$ molecule}

\begin{figure*}[!htbp]
 \centering
\includegraphics[width=0.32\textwidth]{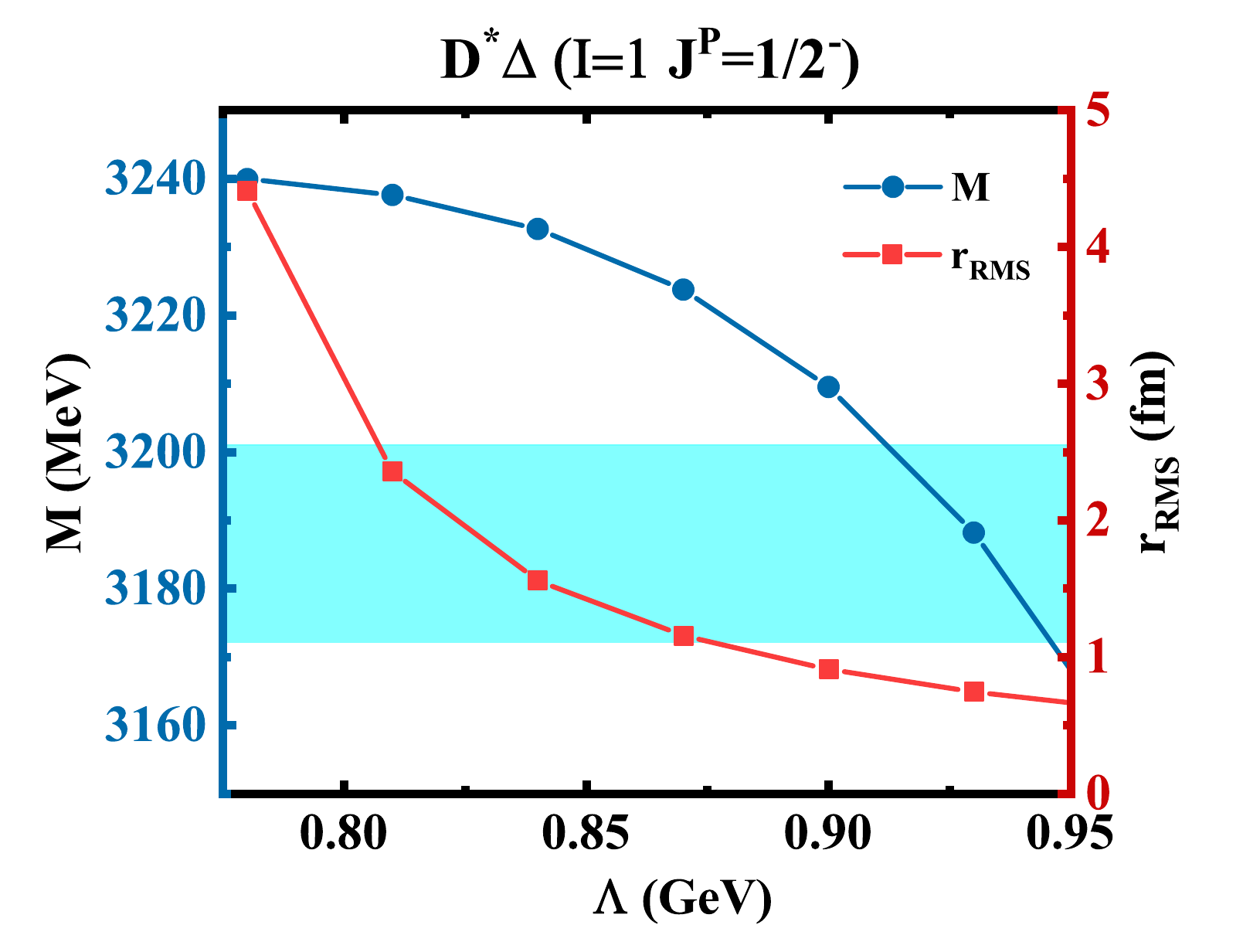}
\includegraphics[width=0.32\textwidth]{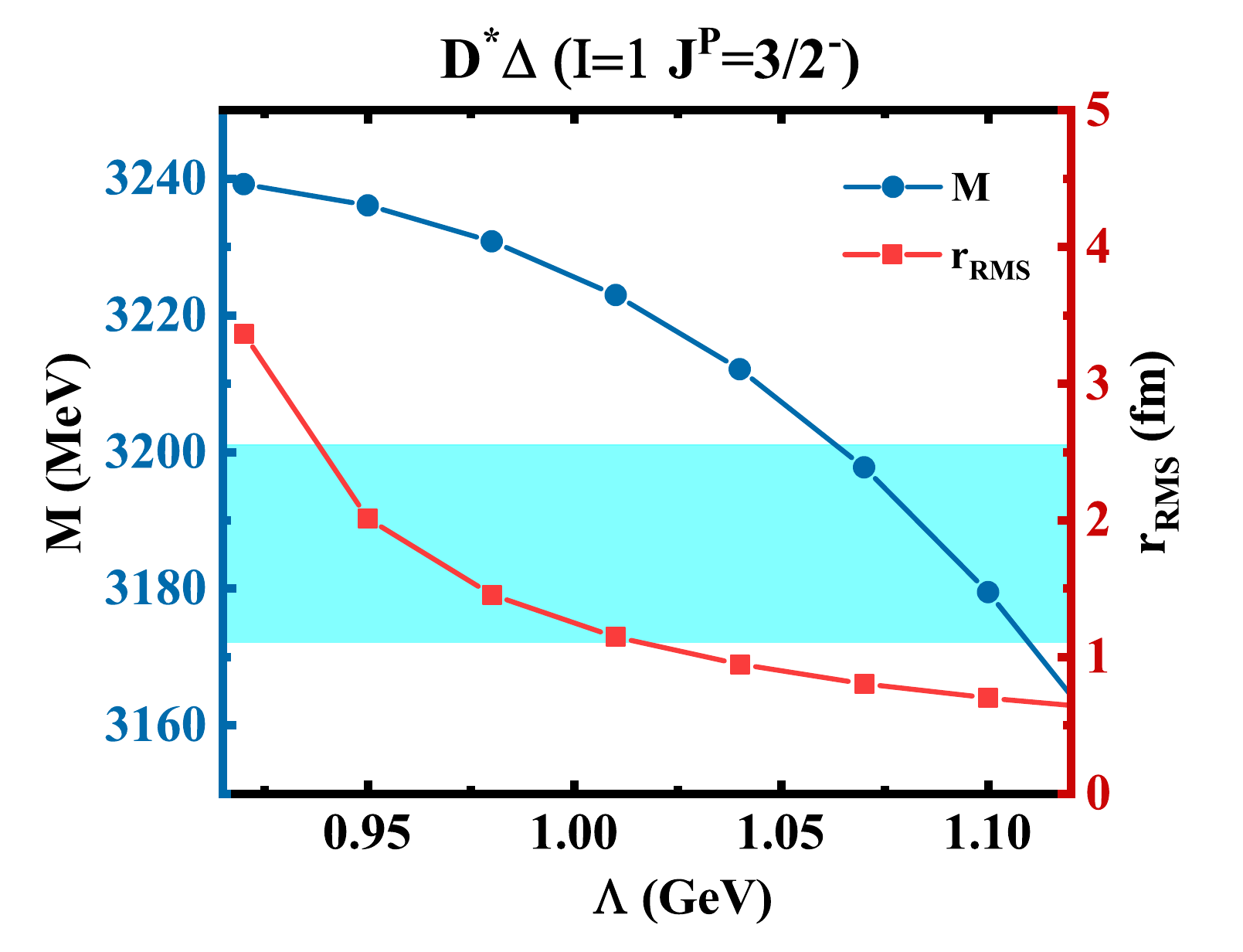}
\includegraphics[width=0.32\textwidth]{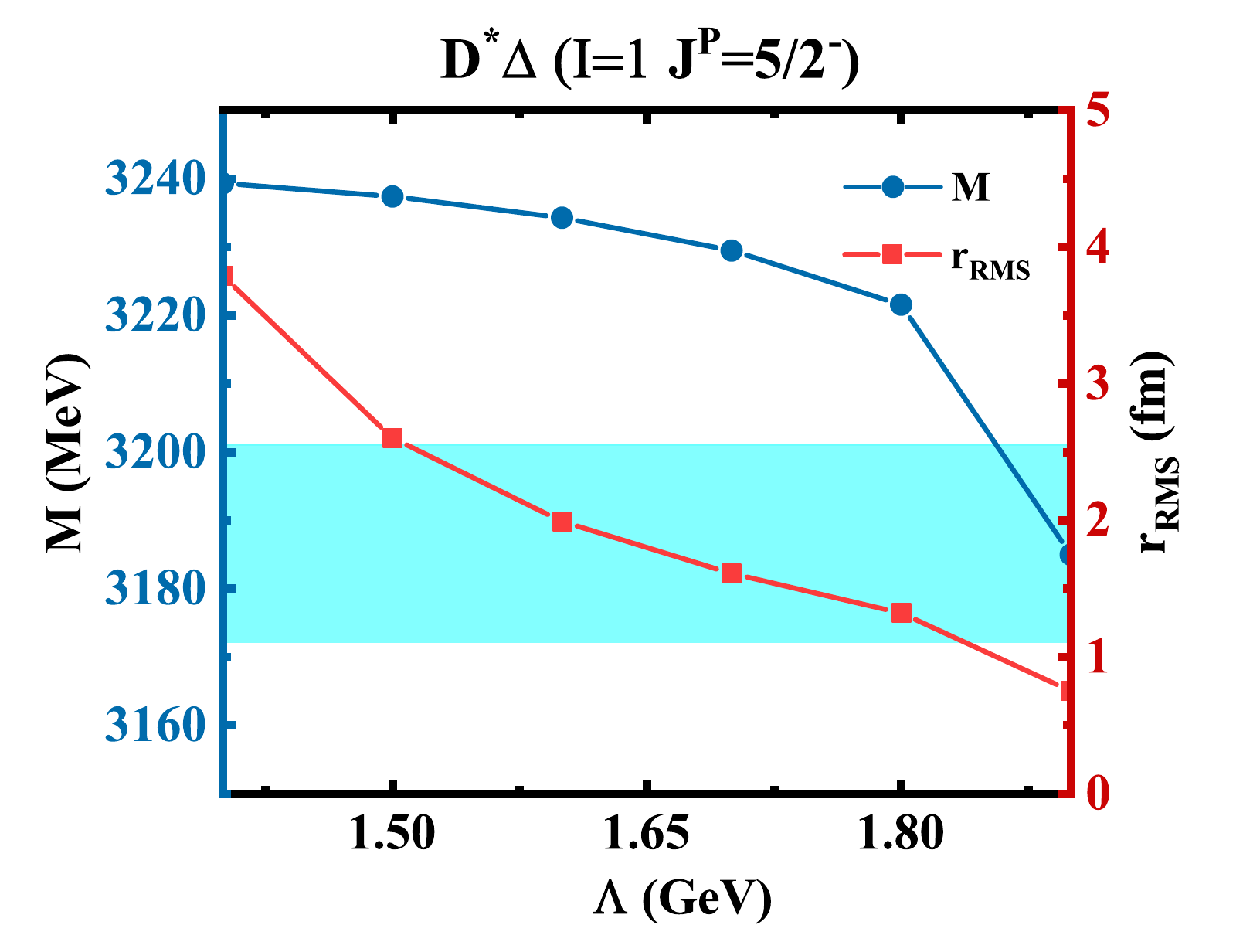}\\
\includegraphics[width=0.32\textwidth]{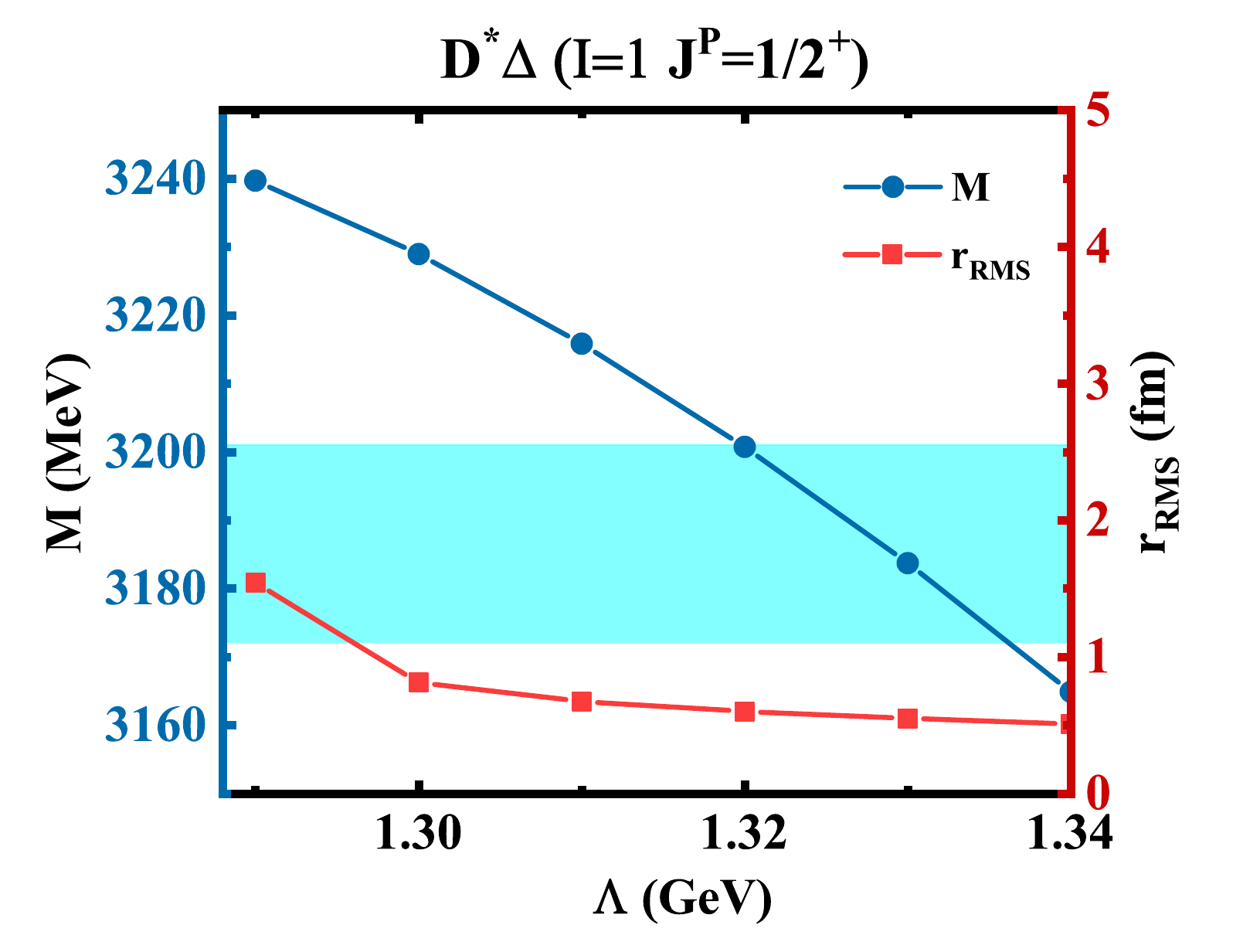}
\includegraphics[width=0.32\textwidth]{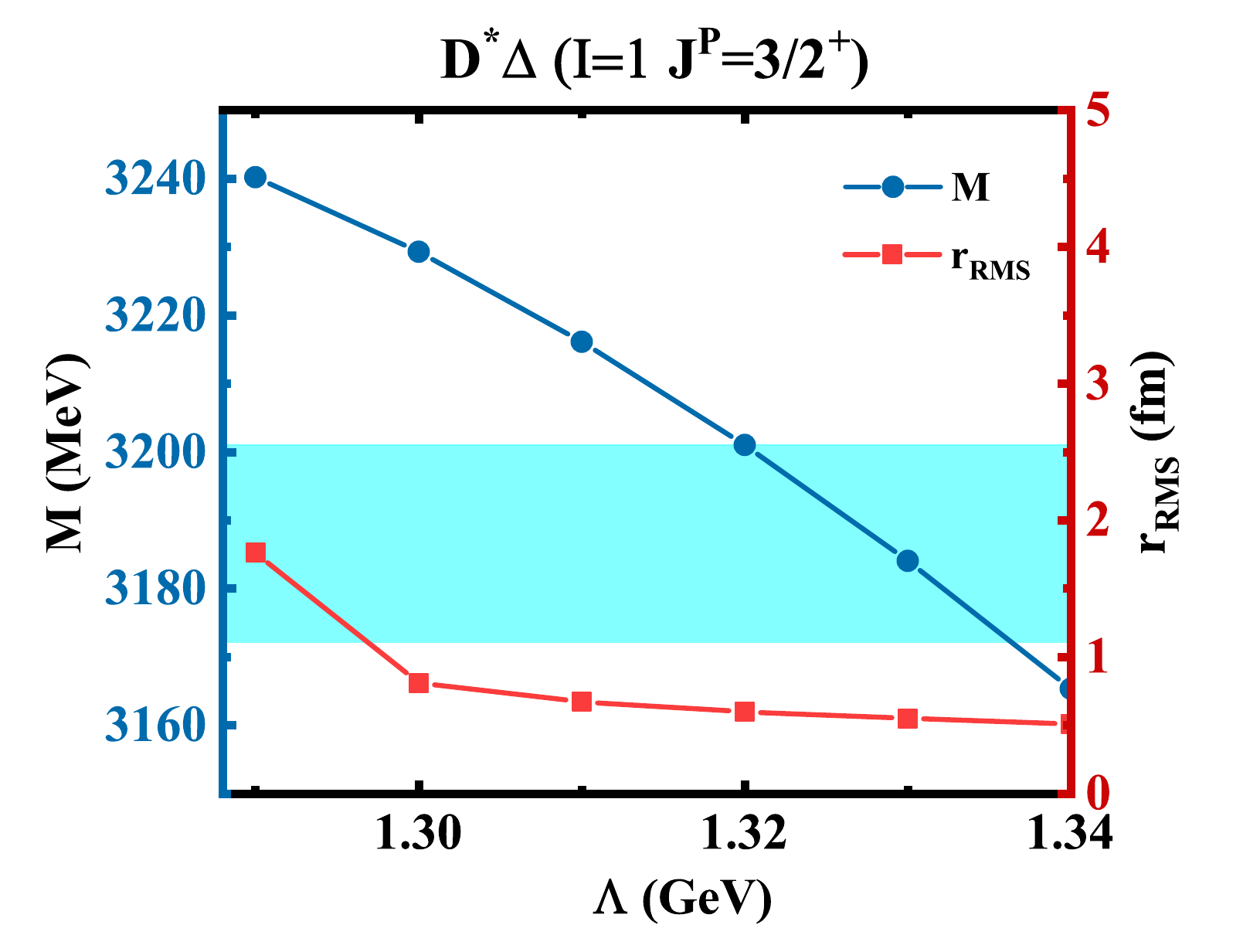}
\includegraphics[width=0.32\textwidth]{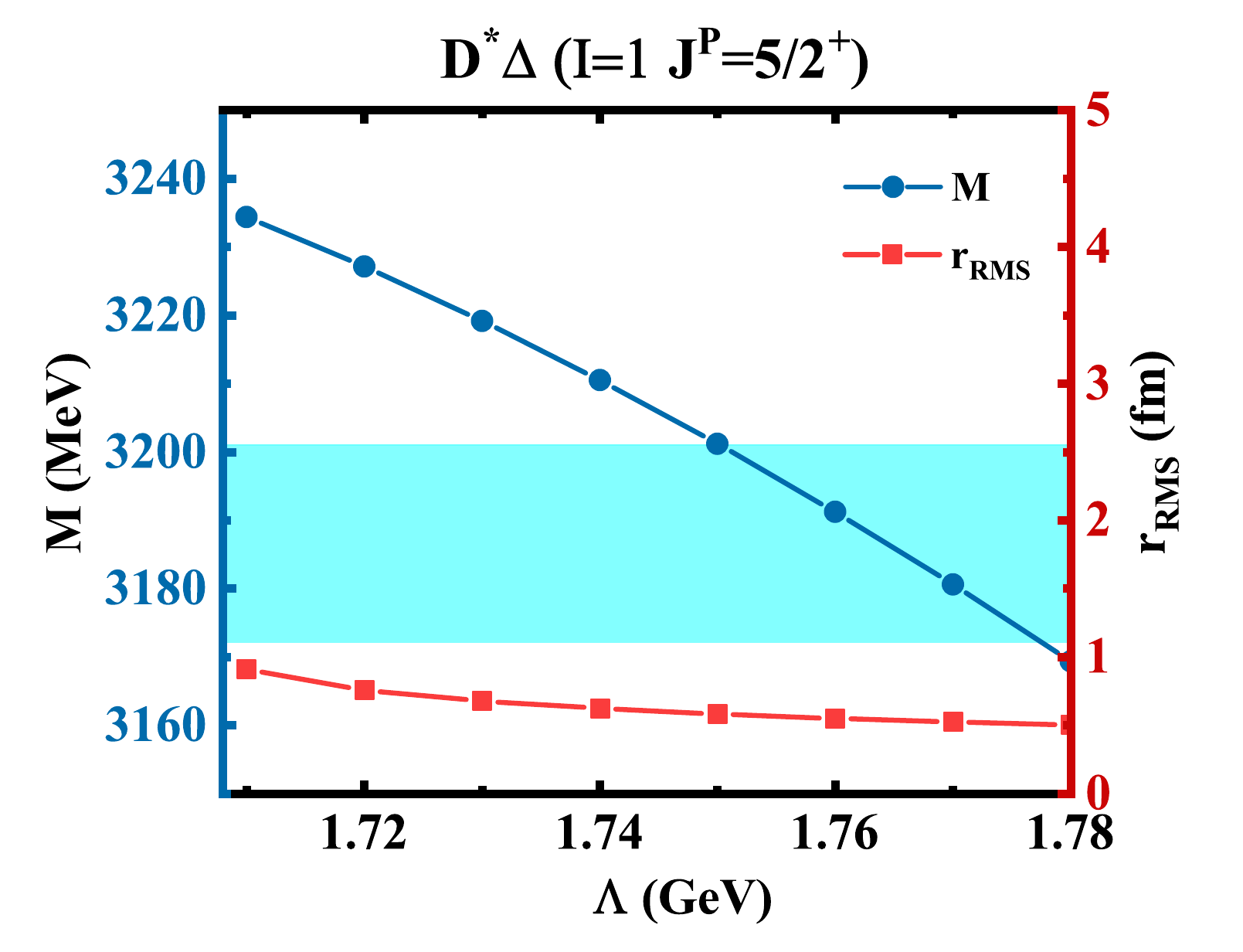}  \\    
    \caption{The cutoff dependence of the obtained masses $M$ (light blue line) and root-mean-square radii $r_{\text{RMS}}$ (red line) for the single $D^*\Delta$ systems with $I(J^P)=1(1/2^{\pm}, 3/2^{\pm}, 5/2^{\pm})$. Here, the shaded region indicates the mass range of \(\Sigma_c(3200)\) after incorporating experimental uncertainties~\cite{LHCb:2026nzu}.}
    \label{fig:example} 
\end{figure*}

The $\Sigma_c(3200)$ state observed in the $\Lambda_c\pi$ final state has a mass of about 3200~MeV and a width exceeding 100~MeV, providing important constraints for identifying its quantum numbers. In the following, we perform a qualitative analysis of each $J^P$ assignment by combining spectroscopic properties and decay behaviours. We first examine the bound state solutions of the $D^*\Delta$ system with isospin $I=1$ in the single channel approximation. The threshold of the $D^*\Delta$ channel is given by $m_{D^*\Delta}^{\rm th}=3240.56$~MeV, so the physical mass is obtained by adding the binding energy $E$ (negative).

\subsubsection{Negative-parity states: $J^P = 1/2^-, 3/2^-, 5/2^-$}
\renewcommand\tabcolsep{0.25cm}
\renewcommand{\arraystretch}{1.7}
\begin{table}[!htpb]
\centering
\caption{The bound state solutions (the binding energy $E$, the root-mean-square radius $r_{RMS}$, and the probabilities for the involved channels) for the $D^* \Delta$ systems with $I=1$. Here, the units for the cutoff $\Lambda$, the binding energy $E$, and the root-mean-square radius $r_{RMS}$ are GeV, MeV, and fm, respectively. Channels that are dominant are marked in bold manner. }\label{dsdelta1}
\begin{tabular}{ccccl}
\toprule[1.5pt]
\toprule[1.5pt] 
 $I(J^P)$  &$\Lambda$    &$E$    &$r_{RMS}$   &$D^* \Delta(^{2}S_{\frac{1}{2}}/^{4}D_{\frac{1}{2}}/^{6}D_{\frac{1}{2}})$  \\ 
 $1(\frac{1}{2}^-)$  &0.78   &$-$0.62&4.41&\textbf{99.40}/0.36/0.24         \\
       &0.82   &$-$4.26&2.02&\textbf{99.11}/0.54/0.35                   \\
         &0.86   &$-$13.33&1.27&\textbf{99.11}/0.55/0.34                   \\\hline

$I(J^P)$  &$\Lambda$    &$E$    &$r_{RMS}$   &$D^* \Delta(^{4}S_{\frac{3}{2}}/^{2}D_{\frac{3}{2}}/^{4}D_{\frac{3}{2}}/^{6}D_{\frac{3}{2}})$   \\
 $1(\frac{3}{2}^-)$   &0.90   &$-$0.30&5.38&\textbf{99.35}/0.18/0.42/0.05         \\
         &0.96   &$-$5.96&1.78&\textbf{98.83}/0.32/0.76/0.09                  \\
    
         &1.02   &$-$20.85&1.07&\textbf{99.02}/0.28/0.63/0.07                  \\\hline

$I(J^P)$  &$\Lambda$    &$E$    &$r_{RMS}$   &$D^* \Delta(^{6}S_{\frac{5}{2}}/^{2}D_{\frac{5}{2}}/^{4}D_{\frac{5}{2}}/^{6}D_{\frac{5}{2}})$   \\
    $1(\frac{5}{2}^-)$ &1.30 &$-$0.22&5.94&\textbf{99.30}/0.05/0.02/0.63         \\
    
         &1.50   &$-$3.15&2.60&\textbf{99.04}/0.06/0.03/0.87                  \\
       
         &1.70   &$-$11.08&1.61&\textbf{99.30}/0.13/0.02/0.54                  \\
\hline

 $I(J^P)$ &$\Lambda$    &$E$    &$r_{RMS}$    &$D^* \Delta(^{2}P_{\frac{1}{2}}/^{4}P_{\frac{1}{2}})$\\
 $1(\frac{1}{2}^+)$          &1.29   &$-$0.82&1.54&\textbf{97.95}/2.05         \\
       &1.30   &$-$11.58&0.81&\textbf{99.19}/0.81                   \\
         &1.31   &$-$24.69&0.68&\textbf{99.62}/0.38                  \\\hline

$I(J^P)$&$\Lambda$      &$E$    &$r_{RMS}$   &$D^* \Delta(^{2}P_{\frac{3}{2}}/^{4}P_{\frac{3}{2}}/^{6}P_{\frac{3}{2}})$\\
$1(\frac{3}{2}^+)$  &1.29   &$-$0.36&1.76&\textbf{98.64}/0.39/0.97         \\
       &1.30   &$-$11.24&0.81&\textbf{99.51}/0.12/0.37                   \\
         &1.31   &$-$24.42&0.67&\textbf{99.76}/0.05/0.19                  \\\hline

$I(J^P)$  &$\Lambda$    &$E$    &$r_{RMS}$    &$D^* \Delta(^{4}P_{\frac{5}{2}}/^{6}P_{\frac{5}{2}})$\\
    $1(\frac{5}{2}^+)$     &1.71   &$-$6.15&0.91&\textbf{99.89}/0.11         \\
       &1.72   &$-$13.37&0.76&\textbf{99.89}/0.11                   \\
         &1.73   &$-$21.35&0.68&\textbf{99.89}/0.11                  \\\hline

  $I(J^P)$  &$\Lambda$    &$E$    &$r_{RMS}$    &$D^* \Delta(^{6}P_{\frac{7}{2}})$\\
 $1(\frac{7}{2}^+)$  & $\times$ & $\times$ & $\times$ & $\times$ \\
\bottomrule[1.5pt]
\bottomrule[1.5pt]
\end{tabular}
\end{table}

As shown in Table~\ref{dsdelta1} and Fig. \ref{fig:example}, all negative-parity states exhibit very high $S$-wave probabilities ($>99\%$). For the $J^P = 1/2^-$ system, a bound solution emerges at $\Lambda = 0.78$~GeV, with $E = -0.62$~MeV and $r_{\mathrm{RMS}} = 4.41$~fm, displaying the typical features of a weakly bound and large-size state. When $\Lambda$ increases to 0.91~GeV, the binding energy deepens to $E=-37.26$~MeV, the mass becomes very close to the $\Sigma_c(3200)$. The cutoff parameter for this state is extremely close to 1~GeV, making it an ideal loosely bound molecular candidate.

For the $J^P = 3/2^-$ system, binding sets in at $\Lambda=0.90$~GeV with $E=-0.30$~MeV and $r_{\mathrm{RMS}}=5.38$~fm; at $\Lambda=1.06$~GeV, $E$ reaches $-37.60$~MeV. The ${}^4S_{3/2}$ wave remains above 95\% throughout. The cutoff parameter here is slightly larger than that in the $I(J^P)=1(1/2^-)$ case, so the OBE interactions can be a little weaker attractive.

For the $J^P = 5/2^-$ system, we find a bound solution only as the cutoff is increased further to $\Lambda=1.30$~GeV, with $E=-0.22$~MeV and $r_{\mathrm{RMS}}=5.94$~fm. When the cutoff increases to $1.89$ GeV, one can reproduce the mass of the $\Sigma_c(3200)$. This cutoff value is obvious larger than those in the $I(J^P)=1(1/2^-)$ and $1(3/2^-)$ cases. Thus, with the same binding energies, the OBE interactions can be much weaker.

In summary, our single-channel calculations show that in the $I=1$ $D^*\Delta$ system, the $S$-wave interaction provides sufficient attraction, and for a fixed cutoff, the attractive strength decreases monotonically with increasing total angular momentum $J$:
$V_{\mathrm{att.}}^{(1/2)} > V_{\mathrm{att.}}^{(3/2)} > V_{\mathrm{att.}}^{(5/2)}$. Within the quantum mechanical two-body scattering framework, the binding energy $E_B$ is positively correlated with the effective interaction strength. Consequently, the above ordering of attraction directly determines the mass hierarchy of the three molecular states: $M_{1/2} < M_{3/2} < M_{5/2}$. The $J=1/2$ state has the largest binding energy (largest downward mass shift), whereas the $J=5/2$ state forms only an extremely shallow bound state, with a mass very close to the $D^*\Delta$ threshold (about $3240$~MeV).

Beyond masses, the strong decay width serves as another key discriminant among different $J^P$ assignments. The observed channel of $\Sigma_c(3200)$ is $\Lambda_c\pi$. For an $S$-wave $D^*\Delta$ molecule, the final-state orbital angular momentum $L_{\Lambda_c\pi}$ in the decay to $\Lambda_c\pi$ is constrained by total angular momentum conservation: the $1/2^-$ state decays via an $S$-wave, whereas the $3/2^-$ and $5/2^-$ states decay via $D$-waves. Given the large kinetic release of about 900~MeV in the $\Lambda_c\pi$ final state, the $D$-wave decay is strongly suppressed by the centrifugal barrier, leading to widths much smaller than that of the $S$-wave decay.

If $\Sigma_c(3200)$ were assigned to the $J=3/2^-$ or $J=5/2^-$ state, then under the ordering $V_{\mathrm{att.}}^{(1/2)} > V_{\mathrm{att.}}^{(3/2)} > V_{\mathrm{att.}}^{(5/2)}$, the $J=1/2$ channel would necessarily form a more deeply bound and lower-mass state than the $J=3/2$ one. If $\Sigma_c(3200)$ were the $J=3/2$ state, the $J=1/2$ state would have a mass below 3200~MeV and would leave a more prominent signal in the $\Lambda_c\pi$ spectrum, yet no such structure has been observed experimentally. This ``missing ground-state member'' constitutes an irreconcilable contradiction for this assignment. The $J=5/2$ interpretation suffers from the same issue of a missing lower-lying state.

In contrast, assigning $\Sigma_c(3200)$ as the $J=1/2^-$ state naturally avoids all these contradictions: it naturally has the largest binding energy, quantitatively explaining the mass position near 3200~MeV; its $S$-wave decay naturally accounts for the significant experimental signal in the $\Lambda_c\pi$ mass spectrum; and it predicts that the $J=3/2^-$ and $J=5/2^-$ states lie near the $D^*\Delta$ threshold, with $\Lambda_c\pi$ not being their dominant decay mode. Instead, channels such as $D^*N$, $DN\pi$, $\Lambda_c\rho$, $\Lambda_c\pi\pi$, and $\Sigma_c\pi\pi$ may be their important strong decay modes.

It is worth noting that the BaBar Collaboration previously reported the $X_c(3250)$ structure in the $\Sigma_c^{++}\pi^-\pi^-$ invariant mass spectrum, with $M=3245\pm20$~MeV and $\Gamma=108\pm60$~MeV \cite{BaBar:2012ekl}. Its mass is very close to the $D^*\Delta$ threshold, indicating a very small binding energy. This feature is naturally consistent with the expectations for the $J=3/2^-$ or $J=5/2^-$ states in our framework. Therefore, we conjecture that $\Sigma_c(3200)$ and $X_c(3250)$ (or some near-threshold structure) may belong to the same $I=1$ $D^*\Delta$ molecular multiplet: $\Sigma_c(3200)$ corresponds to the $J=1/2^-$ state, $X_c(3250)$ corresponds to the $J=3/2^-$ state, and the $J=5/2^-$ state awaits discovery with higher-statistics experiments.

\subsubsection{Positive-parity states: $J^P = 1/2^+, 3/2^+, 5/2^+, 7/2^+$}

As shown in the Table~\ref{dsdelta1} and Fig. \ref{fig:example}, when the cutoff is near 1~GeV, we can find bound state solutions for the $D^*\Delta$ systems with $I(J^P)=1(1/2^+, 3/2^+, 5/2^+, 7/2^+)$. Compared with the negative-parity systemS, the positive-parity states exhibit markedly different behaviour. On one hand, the cutoff values required for binding are somewhat larger, mainly because the positive-parity system does not involve an $S$-wave but is instead composed of $P$-waves with different spins. The presence of the centrifugal barrier weakens the formation of bound states, so that for a given binding energy, a larger cutoff is needed. On the other hand, the coupling among $P$-wave components is more pronounced, especially for lower angular-momentum states. Furthermore, the decay of the positive-parity states $I(J^P)=1(1/2^+, 3/2^+)$ to $\Lambda_c\pi$ proceeds via a $P$-wave ($L=1$), because the $S$-wave final state has negative parity, mismatching the positive initial parity, while the $P$-wave has positive parity and satisfies the parity requirement. The $P$-wave is suppressed by the centrifugal barrier, resulting in relatively small widths.

Consequently, if $\Sigma_c(3200)$ were assigned as a positive-parity $D^*\Delta$ molecule with isospin 1, a serious spectroscopic self-consistency problem arises: the negative-parity states possess stronger attraction, especially the $1/2^-$ state, which would leave a more prominent signal in the $\Lambda_c\pi$ spectrum, yet no such structure has been observed.

In summary, within our model, the optimal quantum-number assignment for $\Sigma_c(3200)$ is $I(J^P)=1(1/2^-)$. Other more shallowly bound $D^*\Delta$ molecular states are expected to exist near the $D^*\Delta$ threshold (around $3240$~MeV). Measurement of the final-state angular distribution is the most direct way to distinguish among different $J$ assignments. Future amplitude analyses of the $\Lambda_c\pi$ decay angular distribution by LHCb and Belle~II should be able to definitively discriminate among the $J=1/2$, $3/2$, $5/2$, and $7/2$  possibilities.

\subsection{Predictions of other $D^{(*)}\Delta$ molecular candidates}

\subsubsection{The $D^*\Delta$ systems with $I=2$}

\renewcommand\tabcolsep{0.25cm}
\renewcommand{\arraystretch}{1.7}
\begin{table}[!htpb]
\centering
\caption{The bound state solutions (the binding energy $E$, the root-mean-square radius $r_{RMS}$, and the probabilities for the involved channels) for the $D^* \Delta$ systems with $I=2$. Here, the units for the cutoff $\Lambda$, the binding energy $E$, and the root-mean-square radius $r_{RMS}$ are GeV, MeV, and fm, respectively. Channels that are dominant are marked in bold manner. }\label{dsdelta2}

\begin{tabular}{ccccl}
\toprule[1.5pt]
\toprule[1.5pt]
$I(J^P)$  &$\Lambda$    &$E$    &$r_{RMS}$   &$D^* \Delta(^{2}S_{\frac{1}{2}}/^{4}D_{\frac{1}{2}}/^{6}D_{\frac{1}{2}})$   \\ 
 $2(\frac{1}{2}^-)$  &1.20   &$-$0.40&5.19&\textbf{98.87}/0.59/0.54         \\
      
         &1.26   &$-$4.31&2.10&\textbf{96.82}/1.62/1.56                  \\
        
         &1.32   &$-$14.35&1.27&\textbf{95.02}/2.49/2.49                  \\\hline

$I(J^P)$  &$\Lambda$    &$E$    &$r_{RMS}$   &$D^* \Delta(^{4}S_{\frac{3}{2}}/^{2}D_{\frac{3}{2}}/^{4}D_{\frac{3}{2}}/^{6}D_{\frac{3}{2}})$   \\
 $2(\frac{3}{2}^-)$   &1.20   &$-$2.96&2.46&\textbf{98.27}/0.56/1.03/0.14         \\
       &1.23   &$-$5.31&1.92&\textbf{97.80}/0.72/1.31/0.18                   \\
         &1.26   &$-$8.44&1.59&\textbf{97.35}/0.87/1.57/0.22                  \\\hline
           
$I(J^P)$  &$\Lambda$    &$E$    &$r_{RMS}$   &$D^* \Delta(^{6}S_{\frac{5}{2}}/^{2}D_{\frac{5}{2}}/^{4}D/^{6}D_{\frac{5}{2}})$    \\
    $2(\frac{5}{2}^-)$  &0.95   &$-$0.04&6.41&\textbf{99.60}/0.05/0.01/0.33         \\
         &1.05   &$-$4.10&2.12&\textbf{98.55}/0.19/0.05/1.21                  \\
         &1.15   &$-$12.56&1.37&\textbf{97.62}/0.31/0.09/1.98                  \\ \hline

 $I(J^P)$&$\Lambda$    &$E$    &$r_{RMS}$    &$D^* \Delta(^{2}P_{\frac{1}{2}}/^{4}P_{\frac{1}{2}})$\\ 
 $2(\frac{1}{2}^+)$           &1.55   &$-$1.04&2.17&25.14/\textbf{74.86}         \\
       &1.58   &$-$7.16&1.32&25.17/\textbf{74.83}                   \\
         &1.61   &$-$15.23&1.08&25.34/\textbf{74.66}                  \\\hline

  $I(J^P)$&$\Lambda$    &$E$    &$r_{RMS}$    &$D^* \Delta(^{2}P_{\frac{3}{2}}/^{4}P_{\frac{3}{2}}/^{6}P_{\frac{3}{2}})$\\
       $2(\frac{3}{2}^+)$  &1.55   &$-$0.72&2.33&14.09/2.92/\textbf{82.99}         \\
       &1.58   &$-$5.28&1.44&14.07/2.91/\textbf{83.02}                   \\
         &1.61   &$-$10.98&1.20&14.25/2.93/\textbf{82.82}                  \\\hline

$I(J^P)$  &$\Lambda$    &$E$    &$r_{RMS}$    &$D^* \Delta(^{4}P_{\frac{5}{2}}/^{6}P_{\frac{5}{2}})$\\
    $2(\frac{5}{2}^+)$     & $\times$ & $\times$ & $\times$ & $\times$ \\\hline

  $I(J^P)$  &$\Lambda$    &$E$    &$r_{RMS}$    &$D^* \Delta(^{6}P_{\frac{7}{2}})$\\
 $2(\frac{7}{2}^+)$ & $\times$ & $\times$ & $\times$ & $\times$ \\
\bottomrule[1.5pt]
\bottomrule[1.5pt]
\end{tabular}
\end{table}

We now systematically assess the viability of forming loosely bound molecular states in the $D^*\Delta$ system with isospin $I=2$. The results for each quantum-number channel are presented in Table \ref{dsdelta2}. We can obtain loosely bound states solutions for the single $D^*\Delta$ systems with $I(J^P)=2(1/2^-)$, $2(3/2^-)$, and $2(5/2^-)$ as the cutoff values are taken around 1.00 GeV, and the $S-$wave components are the dominant channels with their probabilities exceeded $95\%$ with the binding energies larger than $-10$ MeV. In addition, we find the attractive strength increases with the increasing of total angular momentum in the $I=2$ cases, i.e., $V_{\mathrm{att.}}^{(1/2)} < V_{\mathrm{att.}}^{(3/2)} < V_{\mathrm{att.}}^{(5/2)}$.

For the single $D^*\Delta$ systems with $I(J^P)=2(1/2^+, 3/2^+)$, in the cutoff region of $\Lambda>1.55$ GeV, we can obtain loosely bound states solutions, which match the characteristics of loosely bound molecules. The dominant channels are ${}^4P_{1/2}$ and ${}^6P_{3/2}$, respectively. These larger cutoff values than those in the negative parity cases reduce the likelihood of being candidates for hadronic molecular states.

For the single $D^*\Delta$ systems with $I(J^P)=2(5/2^+, 7/2^+)$, within the scanned $\Lambda$ range, we find no bound state solutions for such channels. This indicates that the $P$-wave interaction between the $D^*$ and $\Delta$ is insufficient to provide the necessary attraction for molecular formation at high spins.

In summary, we can predict the existence of another five $D^*\Delta$ molecular candidates with $I(J^P)=2(1/2^{\pm}, 3/2^{\pm}, 5/2^-)$.

\subsubsection{The $D\Delta$ systems with $I=1, 2$}

\renewcommand\tabcolsep{0.41cm}
\renewcommand{\arraystretch}{1.5}
\begin{table}[!htpb]
\centering
\caption{The bound state solutions (the binding energy $E$, the root-mean-square radius $r_{RMS}$, and the probabilities for the involved channels) for the single $D\Delta$ systems. Here, the units for the cutoff $\Lambda$, the binding energy $E$, and the root-mean-square radius $r_{RMS}$ are GeV, MeV, and fm, respectively. Channels that are dominant are marked in bold manner. }\label{D1L1}
\begin{tabular}{cccclll}
\toprule[1.5pt]
\toprule[1.5pt]

$  I(J^P) $  &$\Lambda$    &$E$    &$r_{RMS}$      &$D\Delta(^{4}S_{\frac{3}{2}}/^{4}D_{\frac{3}{2}})$    \\\hline 
$1(\frac{3}{2}^-)$     
  & 1.10 & -0.77 & 4.28 & \textbf{100.00}/0.00 \\
  & 1.16 & -3.84 & 2.21 & \textbf{100.00}/0.00 \\
  & 1.22 & -9.00 & 1.55 & \textbf{100.00}/0.00 \\
$2(\frac{3}{2}^-)$             & 1.20 & -1.38 & 3.41 & \textbf{100.00}/0.00 \\
 & 1.30 & -5.95 & 1.83 & \textbf{100.00}/0.00 \\
  & 1.40 & -13.01 & 1.33 & \textbf{100.00}/0.00 \\\bottomrule[1.5pt]
\bottomrule[1.5pt]
\end{tabular}
\end{table}

We then search for possible bound state solutions for the single $D\Delta$ systems with $I(J^P)=1,2(1/2^+, 3/2^{\pm}, 5/2^+)$, when we vary cutoff value in the region of $\Lambda\leq 2.50$ GeV. The corresponding bound states solutions are collected in Table~\ref{D1L1} and \ref{D1L2}. We firstly perform a single $D\Delta$ channel analysis. Our results show there can emerge loosely bound states solutions for the single $D\Delta$ systems with $I(J^P)=1(3/2^-)$ and $2(3/2^-)$ with the cutoff value around 1.00 GeV. After considering the coupled channel effects, we can obtain loosely bound states solutions with smaller cutoff values for the coupled $D\Delta/D^*\Delta$ states with $I(J^P)=1(3/2^-)$ and $2(3/2^-)$, the wave function is dominated by the \(D\Delta({}^{4}S_{3/2})\) component, with the probabilities over \(85\%\), and the RMS radii are larger than 1.00 fm, thus, they can be regarded as predominantly \(D\Delta\) molecular candidates. The coupled channel effects play a positive role in the formative to these two bound states. 

Here, we also explore the existence of possible $D\Delta$ molecular candidates with positive parity. In the single channel cases, when we vary cutoff value in the region of $\Lambda\leq 2.50$ GeV, we cannot obtain bound state solutions. We further investigate the impact of coupled channel effects between the $D\Delta$ and $D^*\Delta$ systems on the bound state properties. As shown in Table \ref{D1L2}, bound states solutions can emerge around $\Lambda$ around 1.35 GeV for the systems with $I=1$, since their binding energies depend strongly on the choice of the cutoff parameter, these bound states cannot be recommended as good molecular candidates. For the $I=2$ systems, the OBE interactions become weaker attractive, in particular, for the system with $J^P=5/2^+$, we cannot find bound state solutions in the cutoff region $\Lambda<2.50$ GeV. 

However, for the coupled $D\Delta/D^*\Delta$ states with $I(J^P)=2(1/2^+, 3/2^+)$, their bound states solutions are consistent with the characteristics of shallowly bound hadronic molecular states; that is, when the binding energy is in the range of a few to a dozen MeV, the RMS radii are around 1 fm or larger. Therefore, these two can also be regarded as candidates for hadronic molecular states, and their dominant channels are both $D\Delta$.

In summary, the $I(J^P)=1(3/2^-)$ and $2(1/2^+, 3/2^{\pm})$ states in the shallow-binding region exhibit the typical features of loose molecular states: reasonable cutoff values, radii of a few fm, and $D\Delta$ dominated wave functions. All positive-parity states with $I=1$, by contrast, cannot be consistently interpreted as loose $D^{(*)}\Delta$ molecules. Nevertheless, the OBE effective potentials in these systems do provide sufficient attraction in certain channels, and it is worthwhile to further explore the possible existence of resonant states beyond the bound region.

\renewcommand\tabcolsep{0.07cm}
\renewcommand{\arraystretch}{1.5}
\begin{table}[!htpb]
\centering
\caption{The bound state solutions (the binding energy $E$, the root-mean-square radius $r_{RMS}$, and the probabilities for the involved channels) for the coupled $D\Delta/D^* \Delta$ systems. Here, the units for the cutoff $\Lambda$, the binding energy $E$, and the root-mean-square radius $r_{RMS}$ are GeV, MeV, and fm, respectively. Channels that are dominant are marked in bold manner. The masses thresholds for the involved channels are  $m_{D\Delta}=3099.25$ MeV and $m_{D^*\Delta}=3240.56$ MeV, respectively.}\label{D1L2}
\begin{tabular}{cccclll}
\toprule[1.5pt]
\toprule[1.5pt]
 $  I(J^P) $  &$\Lambda$    &$E$    &$r_{RMS}$      &$D\Delta(^{4}S_{\frac{3}{2}}/^{4}D_{\frac{3}{2}})$   &$D^*\Delta(^{4}S_{\frac{3}{2}}/^{2}D_{\frac{3}{2}}/^{4}D_{\frac{3}{2}}/^{6}D_{\frac{3}{2}})$ \\ \hline
$1(\frac{3}{2}^-)$     
   & 0.90 & -0.55 & 4.68 & \textbf{97.56}/0.01 & 2.19/0.01/0.06/0.17 \\
  & 0.92 & -3.79 & 2.11 & \textbf{92.41}/0.02 & 7.13/0.01/0.09/0.34 \\
  & 0.94 & -11.24 & 1.29 & \textbf{84.97}/0.04 & 14.50/0.01/0.08/0.40 \\
$2(\frac{3}{2}^-)$    & 1.10 & -1.97 & 2.94 & \textbf{98.83}/$\sim$ & 0.55/0.04/0.23/0.35 \\
 & 1.15 & -5.33 & 1.92 & \textbf{98.00}/$\sim$ & 0.88/0.08/0.41/0.63 \\
 & 1.20 & -10.35 & 1.47 & \textbf{97.14}/$\sim$ & 1.15/0.12/0.64/0.95 \\\hline
$  I(J^P) $  &$\Lambda$    &$E$    &$r_{RMS}$      &$D\Delta(^{4}P_{\frac{1}{2}})$   &$D^*\Delta(^{2}P_{\frac{1}{2}}/^{4}P_{\frac{1}{2}})$ \\
$1(\frac{1}{2}^+)$            & 1.35 & -12.17 & 0.70 & \textbf{54.74}&2.11/43.15 \\
 & 1.36 & -30.58 & 0.56 & \textbf{49.39}&3.70/46.91 \\
 & 1.37 & -51.69 & 0.50 & 45.23&5.95/\textbf{48.82} \\
$2(\frac{1}{2}^+)$             & 1.59 & -0.47 & 1.81 & \textbf{59.83}&9.49/30.68 \\
 & 1.60 & -4.61 & 1.11 & \textbf{54.15}&10.91/34.94 \\
 & 1.61 & -9.20 & 0.95 & \textbf{50.86}&11.77/37.37 \\\hline
  
 $  I(J^P)$  &$\Lambda$    &$E$    &$r_{RMS}$      &$D\Delta(^{4}P_{\frac{3}{2}})$   &$D^*\Delta(^{2}P_{\frac{3}{2}}/^{4}P_{\frac{3}{2}}/^{6}P_{\frac{3}{2}})$ \\
 $1(\frac{3}{2}^+)$           & 1.34 & -1.70 & 1.07 & \textbf{61.18}&0.12/38.58/0.12 \\
 & 1.35 & -20.59 & 0.59 & \textbf{54.10}&0.20/45.62/0.08 \\
 & 1.36 & -42.46 & 0.51 & \textbf{50.87}&0.31/48.77/0.05 \\
 $2(\frac{3}{2}^+)$            & 1.59 & -2.01 & 1.33 & \textbf{59.44}&1.64/3.11/35.81 \\
 & 1.60 & -6.49 & 1.03 & \textbf{55.46}&1.87/3.43/39.24 \\
 & 1.61 & -11.38 & 0.91 & \textbf{52.76}&2.05/3.67/41.52 \\\hline

  $  I(J^P)$  &$\Lambda$    &$E$    &$r_{RMS}$      &$D\Delta(^{4}P_{\frac{5}{2}})$   &$D^*\Delta(^{4}P_{\frac{5}{2}}/^{6}P_{\frac{5}{2}})$ \\
   $1(\frac{5}{2}^+)$             & 1.35 & -14.67 & 0.65 & \textbf{55.29}&44.68/0.03 \\
& 1.36 & -34.08 & 0.54 & \textbf{51.41}&48.57/0.02 \\
 & 1.37 & -55.83 & 0.49 & 48.88&\textbf{51.11}/0.01 \\
   $2(\frac{5}{2}^+)$            & $\times$ & $\times$ & $\times$ & $\times$ & $\times$\\
\bottomrule[1.5pt]
\bottomrule[1.5pt]
\end{tabular}
\end{table}

\section{Summary}\label{sec4}

In this work, we have systematically investigated the interactions between the ground-state charmed mesons and the $\Delta$ baryon within the OBE framework, with the aim of elucidating the nature of the newly observed $\Sigma_c(3200)^0$ state by the LHCb Collaboration and predicting possible exotic charmed-baryon molecular candidates. By constructing the relevant wave functions and deriving the OBE effective potentials, we have solved the coupled-channel Schrödinger equations for the $D^{(*)}\Delta$ systems with isospin $I=1$ and $2$, covering a wide range of quantum numbers $J^P = 1/2^\pm, 3/2^\pm, 5/2^\pm, 7/2^+$.

Our results show that for the isovector $D^*\Delta$ systems, the negative-parity $S$-wave states can provide the most consistent interpretation of $\Sigma_c(3200)$. Among all quantum numbers, the $I(J^P)=1(1/2^-)$ assignment can be favoured: it can form a loosely bound state with a cutoff near $1$~GeV, yielding a mass around $3200$~MeV and a size typical of hadronic molecules. Its $S$-wave decay to $\Lambda_c\pi$ can naturally explain the observed signal, whereas the $J^P=3/2^-$ and $5/2^-$ partners, which decay via $D$-waves, can be predicted to lie closer to the $D^*\Delta$ threshold with suppressed widths. Assigning $\Sigma_c(3200)$ to higher $J$ would imply a deeper unobserved $J^P=1/2^-$ state, contradicting data. Positive-parity states require larger cutoffs and yield smaller radii, making them less compatible with a loose molecular picture.

For the $I=2$ sector, we can predict five molecular candidates with $I(J^P)=2(1/2^\pm,3/2^\pm,5/2^-)$. The negative-parity states are dominated by $S$-wave components and can be bound with cutoff values around $1$ GeV, whereas the positive-parity states require larger cutoffs and are therefore less likely to be interpreted as loosely bound molecules. No bound-state solution is found for $2(5/2^+, 7/2^+)$ in the scanned cutoff region.

After considering the coupled channel effects between $D\Delta$ and $D^*\Delta$, several additional shallowly bound molecular candidates emerge. In particular, the coupled states with $I(J^P)=1,2(3/2^-)$ exhibit typical molecular features. For the positive-parity coupled channels, the $I(J^P)=2(1/2^+, 3/2^+)$ states may also be regarded as possible $D\Delta$-dominated molecular candidates. In contrast, the positive-parity states with $I=1$ yield small radii and large binding energies, falling outside the loose-molecule regime.

Finally, we stress that decisive discrimination of the quantum numbers requires measurement of the $\Lambda_c\pi$ decay angular distribution, which we encourage in future LHCb and Belle~II analyses. The predicted $I=2$ states and near-threshold partners can offer clear experimental targets in final states such as $\Sigma_c^{(*)}\pi\pi$ and $D^{(*)}N$. Further theoretical progress including three-body effects and lattice QCD comparisons will be valuable.

\section*{ACKNOWLEDGMENTS}

This project is supported by the National Natural Science Foundation of China under Grant No. 12305139, and the Xiaoxiang Scholars Programme of Hunan Normal University. F. L. Wang is also supported by the National Natural Science Foundation of China under Grants No. 12405097 and No. 12335001.

\end{document}